\documentstyle[12pt,epsf]{article}
\epsfverbosetrue
\def\citenum#1{{\def\@cite##1##2{##1}\cite{#1}}}
\def\citea#1{\@cite{#1}{}}

\def\Om{\Omega(s,b)}
\def\xo{e^{-2 \Omega(s,b)}}
\def\xoo{e^{- \Omega(s,b)}}

\def\beq{\begin{equation}}
\def\eeq{\end{equation}}
\def\bea{\begin{eqnarray}}
\def\eea{\end{eqnarray}}

\def\bbbz{{\mathchoice {\hbox{$\sf\textstyle Z\kern-0.4em Z$}}
{\hbox{$\sf\textstyle Z\kern-0.4em Z$}}
{\hbox{$\sf\scriptstyle Z\kern-0.3em Z$}}
{\hbox{$\sf\scriptscriptstyle Z\kern-0.2em Z$}}}}
\relax
\begin{document}
%
\begin{titlepage}
\noindent
FNAL 93-000  \hfill  TAUP 2030-93  \\
                       \\[9ex]
\begin{center}
{\Large \bf Large Rapidity Gaps
            in pp Collisions                  }   \\[11ex]

{\large E.\ Gotsman   *                         }    \\[1ex]
{\large E.M.\ Levin   **                           }    \\[1ex]
{\large U.\ Maor    *                             }    \\[1.5ex]
{*  School of Physics and Astronomy  }      \\
{Raymond and Beverly Sackler Faculty of Exact Sciences} \\
{Tel Aviv University, Tel Aviv}  \\ [1.5ex]
 { **     Fermi National Accelerator Laboratory }  \\
{P.O. Box 500, Batavia, Illinois 60510} \\
%
[21ex]
{\large \bf Abstract}
\end{center}
\begin{quotation}
The survival probability of large rapidity gaps in pp collisions is
calculated for several different eikonal models of the Gaussian form.
Results obtained for models based on
partonic interactions are quite similar.  The
Regge-pole model predicts a  higher value of
$ < \vert S \vert^{2} > $ .
\end{quotation}
\end{titlepage}
%
\section {Introduction}
It has recently been suggested \cite{1,2} that the observation of a large
rapidity gap in the $ \eta - \phi $ lego plot, constructed for
 exceedingly high energy p-p interactions, may serve as a signature
for W-W fusion associated with the production of a Higgs boson.
 The practical utilization of this idea as a useful trigger for
 rare electroweak processes depends on ones ability to reliably
assess the survival probability $ < \vert S \vert^{2} > $ . This is
defined \cite {2} as the fraction of events for which spectator events
do not fill the rapidity gap of interest.
\par The survival probability is easily defined in the eikonal
model in impact parameter space.  We use amplitudes normalised
so that
\beq
    \frac{d \sigma}{dt} = \pi \vert f(s,t) \vert ^{2}
\eeq
\beq
   \sigma_{tot} = 4 \pi Im f(s,0)
\eeq
\beq
 a(s,b) = \frac{1}{2 \pi} \int d{\bf q} e^{- i{\bf q.b}}
f(s,t)
\eeq
{}From which we derive the b-space formulae \cite{3} :
 \beq
 \sigma_{tot} = 2 \int d{\bf b} Im a(s,b)
\eeq
 \beq
 \sigma_{el} = \int d{\bf b} \vert a(s,b) \vert^{2}
\eeq
    s-channel unitarity implies that
$ \vert a(s,b) \vert \leq $ 1, and when written in a diagonalized
form we have
 \beq
2 Im a(s,b) = \vert a(s,b) \vert^{2} + G_{in}(s,b)
\eeq
  from which we obtain for the inelastic cross section
 \beq
\sigma_{in} = \int d{\bf b} G_{in}(s,b)
\eeq
 s-channel unitarity is most easily enforced  in the eikonal approach
where, assuming that a(s,b) is purely imaginary, we can write
 \beq
a(s,b) = i ( 1 - e^{- \Omega(s,b)} )
\eeq
 where the eikonal $ \Omega (s,b) $ is a real function.
\par Our assumption that a(s,b) is purely imaginary  is not
compatible with analyticity and crossing symmetry. These are,
easily restored upon substituting
 $ s^{\alpha } \rightarrow  s^{\alpha} e^{- i \frac{\pi \alpha}{2}} $.
 From Eqn. (7) we can express $ G_{in}(s,b) $ as a function of
 $ \Omega(s,b) $
 \beq
G_{in}(s,b) =1 - e^{-2 \Omega(s,b)}
\eeq
 Note that P(s,b) = $ e^{ - 2 \Omega (s,b)}$ is the probability that no
inelastic interaction takes place at impact parameter b.
\par We follow Bjorken \cite{2} and define
 $ < \vert S \vert ^{2} > $ as the normalized multiplication
of two quantities. The first is a convolution over the parton
densities of the two interacting projectiles presenting the cross
section for the hard parton-parton collision under discussion. The
second  P(s,b) is the probabilty that no other interaction takes place
in the rapidity interval of interest.
In the eikonal formalism we have:
\beq
 < |S|^{2} > =  \frac{\int a_{H}(s,b) P(s,b) d^{2}b}
 { \int a_{H}(s,b) d^{2}b}
 \eeq
  where $ a_{H}(s,b) $ denotes the amplitude
    associated with hard collisions that
can be expressed through the eikonal $ \Omega _{H}(s,b) $ using Eqn. (8).
 Some preliminary calculations of $ < \vert S \vert ^{2} > $
 have been presented in Ref.  \cite{2}. Following this
 pioneering effort there were also  a number of attempts made
to estimate $ < \vert S \vert ^{2} > $
  using Monte Carlo techniques \cite{4,5}.
   It is important to note that these assessments of the survival
probability are model dependent. A number of models  are available
[6 - 11] which provide  a good reproduction of the data in the
ISR - Tevatron range. As we shall show these models differ in their
estimates of $ \Omega_{H}(s,b) $ and P(s,b) in the high energy
 limit of LHC and SSC. It is therefore pertinent to carefully check the
  dependence of $ < \vert S \vert^{2} > $ on the phenomenological
 input required in Eqn. (10). This is the main aim of this note,
 where we have attempted a systematic study of $ < \vert S \vert^{2} > $
and it's sensitivity to the input parameters.
 \par In the following we assume that both $ a_{H}(s,b) $
 and $ \Omega(s,b)$ in  P(s,b)
are well approximated by Gaussians
 \beq
     a_{H}(s,b) = \nu_{H}(s) e^{ - \frac{b^{2}}{R^{2}_{H}}}
\eeq
 \beq
   2 \Omega(s,b) = 2 \nu (s) e^{ - \frac{b^{2}}{R^{2}}}
   \eeq
This input  assumption has been verified by the analysis
of Ref.\cite {10,11}, where it has been shown that eikonal models
of this form   provide an excellent reproduction of the  cross
  section data  and particle distributions in the energy range
  $ 5 \leq \sqrt{s} \leq 1800 $ GeV. The main advantage of assuming
the input in a Gaussian form is that the integration in Eqns. (4)
and (10) can be carried out analytically, whence
 the total cross section
\begin{eqnarray}
\sigma_{tot} = 2 \int d{\bf b} ( 1 - \xoo)
  = 2 \pi R^{2}(s) \sum^{\infty}_{n=1}
   \frac{(-1)^{n-1} \nu ^{n}(s)}{n!n}    \nonumber \\
 = 2 \pi R^{2}(s) [ln \nu (s) + C - Ei(- \nu (s) ) ]
\end{eqnarray}
 $  Ei(x) $ denotes the integral exponential function
  $ Ei(x) = \int^{x}_{- \infty} \frac{e^{t}}{t} dt $  ,
and C is the Euler constant ( C  = 0.5773 ).
 For $ \nu(s) \gg $ 1  we have
\beq
\sigma_{tot} \rightarrow 2 \pi R^{2}(s) [ ln \nu(s) + C  - e^{- \nu(s)}]
\eeq
The inelastic cross section is given by
\begin{eqnarray}
\sigma_{in} =  \int d{\bf b} ( 1 - \xo)
 =  \pi R^{2}(s) [ln (2 \nu (s)) + C - Ei(- 2 \nu (s) ) ]
\end{eqnarray}
and therefore   $ \sigma_{el} = \sigma_{tot} - \sigma_{in} $ =
\begin{eqnarray}
 =  \pi R^{2}(s) [ln ( \frac{\nu (s)}{2}) + C +
  Ei(- 2 \nu (s) ) - 2 Ei(- \nu(s) ) ]
\end{eqnarray}
 Again for $ \nu(s) \gg $ 1  we have
\begin{eqnarray}
\sigma_{el} \rightarrow  \pi R^{2}(s) [ ln (\frac{ \nu(s)}{2}) + C
 + e^{ -2 \nu(s)} - 2 e^{ - \nu (s)} ]  \nonumber
\end{eqnarray}
\par One can extract the values for $ R^{2}(s)$ and $ \nu (s) $
 using the expressions for $ \sigma_{tot}$ and $ \sigma_{el} $
 given in Eqns. (13) and (15). As expected at the Tevatron energy
 where the cross sections are known, the spread in values of
 $ R^{2}(s)$ and $ \nu (s)$ for the different models is rather
 small. It is only when the model parameters are extrapolated
to higher energies does the difference become significant, and
allows one to test the theory on which the parametrization
 is based. We shall estimate the importantance of this dependence
 when making predictions for LHC and SSC energies.
\par The integration of Eqn. (10) yields:
\beq
    < \vert S \vert^{2} > = a (\frac{1}{2 \nu})^{a} \gamma(a,2 \nu)
 \eeq
     where $ a = \frac{R^{2}}{R^{2}_{H}} $
   and $ \gamma(a,x) $  denotes the incomplete gamma function
\beq
\gamma(a,x) = \int ^{x}_{0} z^{a - 1} e^{- z} dz
\eeq
For $ \nu(s) \gg $ 1 Eqn. (17) simplifies to \cite{2} :
\beq
 < \vert S \vert^{2} > \,\, = \,\, \frac{a \Gamma (a)}{ \nu (s)^{a}}
  + \frac{a}{\nu (s)} e^{- \nu (s)}
 \eeq
A general numerical mapping of $ < \vert S \vert^{2} > $
as a function of a and $  \nu $ is shown in Figs. 1 and 2.
\par A realistic asssessment of the survival probability is
subject to a considerable ambiguity, due mainly to the following
reasons:
\\
1) There is no clear definition of the hard component of the
b-space scattering amplitude or eikonal. Moreover, it is not clear
whether the growth of $ \sigma_{tot} $ with s is due to the soft or
to the hard sector. Bjorken \cite{2} has suggested that one can estimate
$ R^{2}_{H}$ from the low energy cross section where $ \sigma_{tot} $
(pp) $ \approx $ 40 mb. Hence $ R^{2}_{H} $ is energy independent and
  a $ > $ 1. Implicit in this estimate is the assumption that
 $ \sigma_{tot} $ growth is associated with the soft sector. This point
of view is not common to all models. In particular, models based on
 parton interactions \cite{6,10} associate the growth of $ \sigma_{tot} $
 with the hard or semi-hard sector. Hence $ R^{2}_{H} $ is
 energy dependent  and a $ \approx $ 1. Bjorken \cite{2} based his
 numerical estimates on the minijet model \cite{6}.
 Indeed in this type of model $ R^{2}_{H} $ approaches a constant in
  the high energy limit. Here, the growth in $ \sigma_{tot} $
 is due to gluon-gluon interactions which are semi-hard. Having
no better insight to this problem, we follow the recipe suggested
  by Bjorken, and fix $ R^{2}_{H} $ from the low energy data at
   $ \sqrt{s}  \approx $ 10 GeV.  \\
2) The input information required in Eqn. (17), i.e. a and 2$ \nu $
  is obtained from fits to the rising pp and $ {\bar p} $p
cross sections. This rise is a consequence of both
a process of
  blackening, i.e. a rise of $ \nu(s) $ ,
 and expansion, i.e. an increase of $ R^{2} $ . However, these
two processes compensate each other as is readily seen
in Eqn. (14). The dependence of
$ < \vert S \vert^{2} > $ on these parameters is completely
different as is evident from Eqn. (19). As we have noted,
there are a number of phenomenological models which reproduce
the available data  well in the 10 $ \leq \sqrt{s} \leq $ 1800 GeV energy
 range. These models differ in their
   estimates of a and 2$ \nu $, and we wish to examine the stability
of $ < \vert S \vert^{2} > $ with respect to variations of the
 input parameters a and $ 2 \nu $ , as deduced from the phenomological
models.                \\
3)  We would like to emphasis that all our estimates of
 $ < \vert S \vert^{2} > $ are based on the eikonal approach, i.e.
on the assumption that the representation for the scattering
amplitude given in Eqn. (8), is a valid approximation at high energies.
This form allows us to formulate the expression for the survival
 probability in a simple and transparent manner.
\par In the following we elaborate on these points by discussing
  a few  input models:  Our results are summarized in Table I,
where we list the values for  $ \nu $,
 $ R^{2}_{H}, R^{2}, \sigma_{tot} $
and $ < \vert S \vert^{2} > $ for energy values of the Tevatron,
LHC and SSC.
\subsection {Minijet-parton model}
   In this model \cite{6} the eikonal for the partonic i-j collision
 is given by
\beq
 \Om(s,b) =W_{ij}(b)\sigma^{QCD}_{ij}(s)
\eeq
  where $ \sigma_{ij}^{QCD} \sim s^{J-1} $ . For $ W_{ij} $ ,
the Chou-Yang formulation is assumed \cite{12}
\beq
   W_{ij}(b) = \frac{\mu_{ij}^{2}}{96 \pi} (\mu_{ij} b)^{3}
           K_{3}( \mu_{ij} b)
 \eeq
   Eqn. (21) is reasonably well approximated by a Gaussian. To realize
 this  we have fitted Eqn. (21) numerically, utilizing the parameters
for $ \mu_{ij} $  given in Ref. \cite{6} . To estimate the
 hard component, we follow the method suggested by Bjorken \cite{2}
and fix $ R^{2}_{H} $ from the model predictions at $ \sqrt{s} $
= 10 GeV, where $ \sigma_{tot} \approx $ 40 mb. This is a natural
choice as in this model, we have relatively small changes
 of $ R^{2}(s) $ with energy. Not suprisingly, our estimates of
$ < \vert S \vert^{2} > $ are compatible with those of Refs.
 \cite{2,4}.
\subsection {Regge-pole model}
 An impressive reproduction of the experimental total cross section
data is obtained \cite{7} by utilizing a simple Regge-pole model.
For our high energy analysis we are interested in the super-critical
Pomeron where the amplitude is given by:
\beq
f(s,t) = i C e^{R^{2}_{0}t} s^{ \alpha (t) - 1} sin[0.5 \pi \alpha(t)]
\eeq
with $ \alpha (t) = 1 + \epsilon + \alpha^{\prime } t $ .
 Donnachie and Landshoff fit \cite{7} the data
with values of the parameters C = 21.7 mb
and $ \epsilon $ = 0.0808. In addition we utilize a global fit
\cite{8}
to B, the nuclear slope, and obtain $ R^{2}_{0}$ =5.2 Ge$V^{ -2}$
and $ \alpha^{\prime } $ = 0.2 Ge$V^{ -2}$ .
 The b-space transform of Eqn. (20) is
\begin{eqnarray}
   a(s,b) \,\, =
           i C \frac{s^{\epsilon }}{2 \mid\beta \mid^{2}}
exp[- \frac{R^{2}_{1}\cdot b^{2}}{4 \mid\beta \mid^{2}}] \cdot
[R_{1}^{2} sin( \frac{\pi}{2} \alpha (0) + Z)
-    \frac{ \pi}{2} \alpha^{ \prime} cos( \frac{ \pi}{2} \alpha (0)
 + Z)]
\end{eqnarray}
where
\begin{eqnarray}
R_{1}^{2} = R_{0}^{2} + \alpha^{ \prime} lns       \nonumber \\
 \mid \beta \mid^{2} = R_{1}^{4} + \frac{ \pi^{4} \alpha^{\prime 2}}{4}
 \nonumber    \\
 Z = \frac{ \pi \alpha^{\prime} b ^{2}}{8 \mid\beta \mid^{2}}
\end{eqnarray}
As both $ \epsilon $ and $ \alpha^{\prime } \ll $ 1, Eqn. (21)
 is well approximated  by a Gaussian ( see Eqn. (11) ) with
    $$ \nu(s) = \frac{C R^{2}_{1}}{2 \mid\beta \mid^{2}} s^{ \epsilon} $$
\beq
    R^{2}(s) = \frac{4 \mid\beta \mid^{2}}{R^{2}_{1}(s)}
\eeq
  The model in the form suggested by Donnachie and Landshoff \cite{7}
is not appropriate for fitting data at high energies, as for very
high energy it violates unitarity $ (a(s,b=0) > 1 )$   above
$ \sqrt{s} \approx $ 5 TeV. For our evaluation we use  a very similar
 eikonalized version    suggested by
Cudell and Margolis \cite {9}, with  C = 24 mb,
$ \epsilon $ = 0.093 and $ \alpha^{\prime }$ = 0.25 Ge$V^{ - 2}$.
For the hard sector our choice is less straightforward than before,
as $ R^{2}(s) \sim $ ln s, and there is no obvious way of defining
the hard component. In Table I we present $ R^{2}_{H}$ as the
threshold of Eqn. (25). Note that the results obtained from the
 Regge-pole model are in marked contrast to the other models
 investigated in this note. The difference will be elaborated
 apon later.
\subsection {Lipatov-like Pomeron}
   A simple parametrization for the Lipatov-like Pomeron \cite{13},
has been suggested in Ref.  \cite{10}. The eikonal is given by
\beq
  \Om = \frac{a_{1}s^{a_{2}}}{(ln s)^{a_{3}}} \cdot
    e^{- \frac{b^{2}}{R^{2}(s)}}
\eeq
where the following two  parametrizations of $ R^{2}$(s)
 provide good fits to the data:
 \beq
    R^{2}_{L1} = a_{4} + a_{5} (ln s)^{a_{6}}
 \eeq
 \beq
    R^{2}_{L2} = a_{4} + a_{5} \sqrt{ln s} + a_{6} ln s
 \eeq
  $ a_{i} $ are fitted parameters.
In this model a regular Pomeron with trajectory $ \alpha$(0) = 1,
is appended to the Lipatov-like Pomeron. Again the choice of what to
take for $ R^{2}_{H} $ is ambigous, as $ R^{2}$(s) is energy dependent
 and it's low energy limit is exceedingly small. We adopt an arbitrary
 definition as suggested in Ref. \cite{2} and use $ R^{2}_{H} $ to be
  the value at $ \sqrt{s} $ = 10 GeV,
 where $ \sigma_{tot}  \approx $ 40 mb.
\subsection {Dual parton model}
This is a multi-component model \cite {11} describing soft and semihard
multiparticle production. The eikonal is given by
\beq
    \Om  = \Omega_{S}(s,b) + \Omega_{H}(s,b) - \Omega_{TP}(s,b)
  - \Omega_{L}(s,b)
\eeq
 where the last two terms correspond to the triple Pomeron and loop
 contributions. As $ \Omega_{TP}$(s,b) and $ \Omega_{L}$(s,b)
 are reasonably small ,
 this is effectively a two component model whose parameters are:
\beq
\Omega_{S}(s,b) = \frac{ \sigma _{S}}{ 8 \pi R^{2}_{S}} \cdot
             e^{- \frac{b^{2}}{4  R^{2}_{S}}}
\eeq
 where
\begin{eqnarray}
   \sigma_{S} = C s^{\epsilon}        \nonumber   \\
   R^{2}_{S} = B + \alpha^{\prime} ln s
\end{eqnarray}
 with C = 40.8 mb, $ \epsilon $ = 0.076 and $ \alpha^{\prime} $ = 0.24
 Ge$V^{- 2}$.
\beq
\Omega_{H}(s,b) = \frac{\sigma_{H}}{ 8  \pi R^{2}_{H}} \cdot
             e^{- \frac{b^{2}}{4  R^{2}_{H}}}
\eeq
  where   $ R^{2}_{H}$ = B.
 In this type of model $ R^{2}_{H} $ is defined  as the low
energy threshold limit of $ R^{2}_{S}(s) $  given by Eqns. (30) and (31).
The $ \sigma_{H} $ is calculated in lowest order QCD and is dependent
 on the value taken for the $ p_{t}^{min} $ cutoff. Numerical values
for $ \sigma_{H} $ at different energies are given in \cite{11}.

\section{Conclusions}
Our results are summarized in Table I. The results obtained for the
various partonic models \cite{6}, \cite{10}, \cite{11} for
$ < \vert S \vert^{2} > $ in the LHC-SSC energy range are remarkably
 stable. We note that the models of \cite{10} and \cite{11}, even though
 very different in their construction, yield rather similar input
 parameters, as summarized in Table I. Ref. \cite{6} differs from
the above, in that it has the highest values for $ \nu $. These are
compensated for by having the corresponding lowest values for a,
 producing final results which are similar to those obtained in Refs.
\cite{10} and \cite{11}.
\par  The survival probability obtained from the Regge-pole
 model [8-10] are
considerably higher. On examining the input parameters used in our
 calculation, we find that the difference can be traced to the fact
that the $ \nu $ values associated with the Regge-pole model
 are the smallest. Hence, in order to be compatible with the data the
 model requires relatively large values of $ R^{2}_{H} $ and
 $ R^{2} $ which give rise to high $ < \vert S \vert^{2} > $ .
\par We conclude with a more general comment. Clearly the questionable
  aspect  of a calculation
such as presented here, is the fact that the definition
of the hard component is not unique. We have followed Bjorken's
suggestion \cite{2}, and have fixed $ R^{2}_{H} $ from the low
energy data. It is likely that our estimates for $ R^{2}_{H} $
are on the conservative side, so that in reality one could expect
even higher values for the survival probability than are given
in Table I.
\newpage
  \centering
{{ \large \bf Table I }
 { Parameters and predictions of different models }

  \vspace{.5 cm}
 \begin{tabular}{|c|c|c|c|c|c|c|}   \hline \hline
 Model & $ \sqrt{s}$ & $\nu $(s)  & $ R^{2}_{H} $ & $ R^{2} $ &
  $ \sigma_{tot}$ & $ < \vert S \vert^{2} > $   \\
 &  TeV & & Ge$V^{-2}$ & $ GeV^{-2} $ & mb &  \% \\
\hline
 Minijet $^{(6)}$ & 1.8 & 2.50 & 14.41 & 19.74 & 72 & 13.2  \\
 & 16.0 & 3.90 & 14.41 & 22.60  & 107 & 5.5        \\
 & 40.0 & 4.75 & 14.41 & 23.20 & 121 &  3.8 \\
\hline
Regge $^{(7-9)}$ & 1.8 & 1.11 & 25.41 & 35.80 & 76 & 32.6  \\
 & 16.0 & 1.48 & 25.41 & 40.16 & 102 & 22.1 \\
 & 40.0 & 1.68 & 25.41 & 41.99 & 117 & 18.1 \\
\hline
Lipatov 1  $^{(10)}$ & 1.8 & 1.60 & 15.78 & 25.39 & 75 & 19.7   \\
 & 16.0 & 2.69 & 15.78 & 28.02 & 113 & 8.2 \\
 & 40.0 & 3.44 & 15.78 & 29.37 & 134 & 4.9 \\
\hline
Lipatov 2  $^{(10)}$ & 1.8 & 1.44 & 16.19 & 28.64 & 76 & 20.6  \\
 & 16.0 & 2.24 & 16.19 & 32.66 & 115  & 9.2 \\
 & 40.0 & 2.77 & 16.19 & 34.23 & 137 & 5.8 \\
\hline
Dual parton  $^{(11)}$ & 1.8 & 1.83 & 10.56 & 28.47 & 75 & 9.6  \\
 KMRS[B-2] & 16.0 & 2.23 & 10.56 & 32.67 & 109 & 5.3 \\
 & 40.0 & 2.43 & 10.56 & 34.43  & 124 & 4.0 \\
\hline
\hline
 \end{tabular}
\newpage
{\large \bf Figure captions}
  \vspace{.5 cm}

{ \bf Figure 1: }  Contours of percentage of the survival probability
$ < \vert S \vert^{2} > $  as a function of a and $ \nu $ .

{ \bf Figure 2: } Graph of  log ( \% survival probability ) versus
 a for selected values of $ \nu $ .

\newpage

\end{document}